\documentclass[a4paper,100pt]{article}

\usepackage[utf8]{inputenc}
\title{Dimensional Analysis Approach to Experiments in Z pinch Devices}
\author{Miguel Cárdenas\\Universidad de Playa Ancha\\Av. Playa Ancha 850, Valparaíso, Chile\\e-mail: miguel.cardenas@upla.cl}

\begin{document}
\date{}

\large
\maketitle
\begin{abstract}
The physical behavior of discharges in Z pinch devices can be completely deciphered in terms of only three dimensionless parameters. These parameters can be arranged in a way that draw a surface in $3D$ space. This surface compiles all the accessible information on the macroscopic physical behavior of each possible Z pinch discharge. We analyze the practical problems the drawing of this surface encounters and in view of the situation, we devote the remainder of the article to outline a feasible method for estimating the plasma temperature in Z pinch discharges.
\end{abstract} 

\section{Introduction}
The challenge of controlled thermonuclear fusion through magnetic confinement has a long history \cite{post}\cite{kolb}\cite{glasstone}\cite{rose}\cite{krall}. However, despite the tireless efforts and unquenchable hope placed on it, consistent evidence pointing to the achievement of controlled thermonuclear fusion has not yet been observed in actual practice. Furthermore, the difficulties involved in measuring plasma temperature mean we lack records of what the temperatures in experiments have actually been. 

If the plasma temperature were within the thermonuclear range of interest, the problem would be simplified. In that case, the measurements were based on the detection of fusion neutrons emitted by the hot plasma. The method, although plausible, has the drawback that it is only suitable for large and expensive experiments. In small experiments, the expected emission of fusion neutrons by the hot plasma is correspondingly tiny. So, we know
in advance that the expected fusion neutron radiation from small experiments will become undetectable because that depleted neutron radiation will be masked by the background radiation. In this way, the plasma temperature in small experiments will remain a mystery. 

In this situation we face, in one way or another, a great practical difficulty in making expedient progress in controlled thermonuclear fusion research. However, and being fully aware of everything we have pointed out above, we must reiterate that the unavoidable challenge that motivates controlled thermonuclear fusion research  is still to find the trick to dramatically raising the plasma temperature in experiments. 

Over decades of painstaking research work, various devices have been designed to materially achieve magnetic confinement fusion. Among the original projects still in vogue is the Z pinch device \cite{ryutov}\cite{jackson}\cite{reitz-milford}. This experimental setup is the simplest imaginable, and that is precisely why it is so appealing. Even so, the physics at play in Z pinch experiments is challenging and difficult to grasp. But the crux of the problem is that Z pinch experiments depend on too many control parameters. 

In this work, we will focus exclusively on matters concerning the study of Z pinch experiments, so that the rationale for this article is to introduce a more concise way of addressing the Z pinch problem as well as indicating an expeditious manner to deal with the problems involved in getting the plasma temperature.

To achieve this goal, we begin by identifying the parameters that affect the physics of the discharges in Z pinch experiments. Then, appealing to the Buckingham's $\Pi$ theorem of dimensional analysis, we recognize that the Z pinch problem can be fully understood in terms of only three dimensionless parameters \cite{bridgman}\cite{langhaar}\cite{barenblatt}\cite{gukhman}. Two of these parameters are constructed based on the control parameters of the Z pinch experiment, while the third depends on them. The value that this dependent dimensionless parameter reaches in each particular experiment needs to be measured. If we could carry out the measurements, then the performance of  just a few different Z pinch experiments would shed enough light on the problem of discharges in Z pinch devices as a whole.

Unfortunately, at the moment those measurements are not within our reach. Therefore, for the time being, we present an alternative proposal that combines accessible measurements with the appropriate equations of a mathematical model that accounts for the physics of Z pinch discharges \cite{rosenbluth}\cite{cardenas1}\cite{cardenas2}. In this way, we do reveal, in a first approximation, the physics of discharges in Z pinch devices. 

In short, we have to measure the radius of the plasma column as a function of time during a given discharge. With this information in hand and the help of the model, we can access the plasma temperature at the time of the pinch. Repeating this procedure with a selected set of dimensionless control parameters values provides a general picture of the plasma temperature behavior in Z pinch discharges. This dataset can, in turn, be represented as a surface in a $3D$ cartesian coordinate system. 

So now, regardless of the specific Z pinch experiment we are dealing with, that particular experimental setup corresponds just to a point on the surface in $3D$ space. In other words, each Z pinch experiment corresponds to a point in the all-encompassing surface in $3D$ space.
Incidentally, those points condense all the accessible information about the macoscopic physical behavior of a wide variety of Z pinch experiments. 

We will provide in full detail the method for extracting the macroscopic physical information condensed at any point on the aforementioned all-encompassing surface in $3D$ space.

This article is organized as follows: In section 2, we identify the relevant parameters of the Z pinch problem and indicate how to simplify its treatment. In section 3, we present a protocol for accessing the macroscopic physics of the Z pinch problem experimentally. In section 4, we present a mathematical model whose great advantage is that it allows a closed and once-for-all solution to the Z pinch problem. In section 5, we show how the combination of this model with experiments leads to an easy method to estimate the plasma temperature. In section 6, we present and discuss our results. In section 7, we list our conclusions.

\section{The Z pinch problem and its management}

Clearly, the control parameters of a discharge experiment in a Z pinch device are: 

\begin{itemize}
\item The inner radius of the cylindrical container, $r_0$.
\item The length of the cylinder, \textit{i.e.} the electrodes distance, $ l_0$.
\item The mass density of the gas in the cylinder, $\rho_0$.
\item The atomic mass of this working gas, $m$.
\item The total parasitic inductance of the setup, $L_0$.
\item The total capacity of the bank of capacitors, $C_0$.
\item The initial voltage across the bank of capacitors, $V_0$.
\end{itemize}
Let us highlight that the total number of atoms inside the tube, $N_0$, is obtained from the parameters listed above using the formula

\begin{equation}
N_0=\frac{\pi r_0^2 l_0 \rho_0}{m}.
\end{equation}
Similarly, the electrostatic energy initially stored in the bank of capacitors $E_0$, which is what powers the dynamics of the system, is also computed from the parameters listed above through the expression

\begin{equation}
E_0=\frac{1}{2}C_0V_0^2.
\end{equation}
It is important to clarify here that through this research work, we will assume that the plasma is fully ionized, so that its conductivity is infinite and we will also assume that the ohmic resistance of the whole circuit is negligible. Additionally, it should be noted that the choice of the operating gas will typically be either deuterium or helium, so that correspondingly the atomic mass of the working gas will be either $m$ or $2\,m$. Similarly, the ions charge will be either $q$ or $2\,q$. By the way, both ionic mass and ionic charge are important because they affect the plasma current $I(t)$.

When the system comes into operation, the seven listed control parameters determine its further measurable physical properties, \textit{e.g.} the plasma density, the plasma internal energy, the plasma temperature, etc. Thus, to have a complete picture we need the seven listed control parameters plus an additional dependent parameter. To be specific, when we launch the Z pinch discharge, the working plasma begins to gain internal energy and the amount it gains in a given discharge stage, let us say $U$, depends enterely on the seven listed control parameters. Hence, in this case we assimilate $U$ with the missing extra parameter that depends on the other seven control parameters. This is, $U$ corresponds to the eighth parameter of the Z pinch system.
Therefore, the relevant information about discharges in Z pinch experiments can be summarized as
\begin{equation}
U=U(r_0, l_0, \rho_0, m, L_0, C_0, V_0).
\end{equation}
Naturally, the problem immediately arises that for truly gain a general understanding of how $U$ behaves in these experiments, a countless number of experiments would be needed because $U$ depends on too many parameters. Fortunately, there is a more fundamental and economical way to address this problem. That is, we can resort to the Buckingham's $\Pi$ theorem of dimensional analysis which greatly facilitates the research work \cite{bridgman}\cite{langhaar}\cite{barenblatt}\cite{gukhman}. 

Indeed, the application of the $\Pi$ theorem to the present case leads to the conclusion that the bulk-behavior of the plasma in Z pinch experiments can be addressed based on only $3$ dimensionless parameters because there are $8$ original dimensional parameters (see Eqn.(3))  that involve $5$ physical dimensions, namely mass, length, time, current (or charge) and temperature. Therefore, following the prescription provided by the $\Pi$ theorem, the number of relevant dimensionless parameters for this problem is given by the subtraction $(8-5)=3$. So, the complex Eqn.(3) can be substituted with the much simpler formula

\begin{equation}
\Pi_3=\Pi_3(\Pi_1, \Pi_2)
\end{equation}

Dimensionless parameters are constructed based on the physical variables of the system by applying the $\Pi$ theorem. This theorem, it must be said, allows for several options for the resulting dimensionless $\Pi$ groups. However, the central point is that the solution to our problem requires only three dimensionless parameters. Choosing them is, in a way, a matter of taste. Of course, we are aware that some options may seem more common sense than others.
 In the present case, we will obtain the dimensionless parameters by a method alternative to the $\Pi$ theorem and we will not delve into overinterpretations.

\section{Experimental approach to the Z pinch problem}
Within the scheme outlined above, experimental research on Z pinch discharges might perfectly adopt the following protocol:

\begin{itemize}
\item On the basis of the Z pinch relevant parameters, construct the dimensionless parameters $\Pi_1, \Pi_2$ and $\Pi_3$. Below, we will provide specific expressions for $\Pi_1, \Pi_2$ and $\Pi_3$.

The dimensionless parameters  $\Pi_1$ and $\Pi_2$ will be given in terms of the control parameters of the experiment, so we will get them immediately. Instead, the $\Pi_3$ dimensionless parameter is much trickier to get. It will involve carrying out an experimental measurement. 
\item Conduct a number of experiments that sweep a range of $\Pi_1$ and $\Pi_2$ values.
\item Collect the values attained by the dependent dimensionless parameter $\Pi_3$ from each of these experiments.
\item Using this data, plot the function $\Pi_3=f(\Pi_1, \Pi_2)$  in the cartesian system of coordinates $XYZ$. This is, set:

\begin{equation}  
\Pi_1\to X,
\end{equation}

\begin{equation}
\Pi_2\to Y
\end{equation}
and 

\begin{equation}
\Pi_3\to Z.
\end{equation}
\end{itemize}
The surface in $3D$ space constructed in this way summarizes all accessible knowledge of  measurable macroscopic physical properties of generic Z pinch discharges. In other words, no matter what specific Z pinch experiment you are analyzing, the macroscopic behavior of its working plasma  compiles in the surface in $3D$ space that results from the equation

\begin{equation}
\Pi_3=f(\Pi_1, \Pi_2).
\end{equation}

\section{A model-based solution to the Z pinch problem}
The bulk-motion of the plasma during discharges can be treated within a closed mathematical framework known as the snowplow model. The governing dynamics equations for this model have been derived elsewhere \cite{rosenbluth}\cite{cardenas1}\cite{cardenas2}. 

After proceeding with the normalization of the dynamical variables that appear in the fundamental snowplow equations, we obtained that these equations -thus converted into dimensionless- acquired the appearence

\begin{equation}
\left(\frac{d^2r}{dt^2}\right)=\frac{6r^2 (dr/dt)^2-3\alpha^2 I^2-4\int_0^t (1/r)(dr/dt')^3 dt'}{3r(1-r^2)}
\end{equation}
\\and

\begin{equation}
\left(\frac{dI}{dt}\right)=\frac{1-\int_0^tIdt'+\beta (I/r)(dr/dt)}{1-\beta\ln{(r)}}
\end{equation}
where the dimensionless time $t$ is measured in units of $\sqrt{L_0C_0}$, the dimensionless plasma sheath radius $r$ is measured in units of $r_0$ and the dimensionless current $I$ is measured in units of $\sqrt{C_0/L_0}\,V_0$.

On the other hand, the dimensionless parameters $\alpha^2$ and $\beta$ present in these equations were given, in terms of the dimensional Z pinch control parameters, by the expressions

\begin{equation}
\alpha^2=\left(\frac{\mu_0 C_0^2  V_0^2}{4\pi^2  r_0^4\rho_0 }\right)
\end{equation}
and 
\begin{equation}
\beta=\left(\frac{\mu_0 l_0}{2\pi L_0}\right).
\end{equation}
Since we are particularly interested in studying the thermal behavior of the discharges, we choose the third and final dimensionless parameter that describes the physics of our system as the quotient between internal energy and source energy, namely

\begin{equation}
\gamma=\left(\frac{U}{E_0}\right).
\end{equation}

In this way, we have obtained the three dimensionless $\Pi$ groups expected from the $\Pi$ theorem in an alternative manner.
Let us stress that the set of dimensional parameters $(\alpha^2,\beta, \gamma)$ contains the essential information that fully summarizes the macroscopic physics of Z pinch discharges. By the way, the true relationship among $\alpha^2, \beta$ and $\gamma$ is a matter that only experiments can elucidate.

Now, let us remember that in the context of the snowplow model, we have already shown that  \cite{cardenas1}\cite{cardenas2}

\begin{equation}
\left(\frac{U}{E_0}\right)=\left(\frac{\beta}{\alpha^2}\right)\int_0^{t_p}\left(-\frac{1}{r}\left(\frac{dr}{dt}\right)^3dt\right)
\end{equation}
with $t_p$  corresponding to the time when the first contraction of the current sheath happens, \textit{i.e.} $t_p$ is the time of the pinch. Now well, defining $F(\alpha^2, \beta)$ as

\begin{equation}
F(\alpha^2, \beta)=\int_0^{t_p}\left(-\frac{1}{r}\left(\frac{dr}{dt}\right)^3dt\right)
\end{equation}
and replacing Eqn.(13) and Eqn.(15) in Eqn.(14), we obtain the key equation of the problem, namely

\begin{equation}
\gamma=\left(\frac{\beta}{\alpha^2}\right)F(\alpha^2, \beta)
\end{equation}
\\
where the function $r(t)$ that leads to $F$ comes out from solving numerically Eqn.(9) and Eqn.(10).
Let us emphasize that Eqn.(16) is simply the version that the snowplow model provides of Eqn.(8). Indeed, by setting,

\begin{equation}
\alpha^2\to\Pi_1,
\end{equation}

\begin{equation}
\beta\to\Pi_2
\end{equation}
and

\begin{equation}
\gamma\to\Pi_3
\end{equation}
Eqn.(16) and Eqn.(8) become analogous.

With this data we can also draw an overarching surface in $3D$ space, but this will only be an approximation to the surface in $3D$ space given by Eqn.(8). Indeed, now the values taken by the dimensionless parameter $\gamma$ are not exact but come from simulations with the snowplow model. In contrast, the values attained by the dimensionless parameter $\gamma$ in Eqn.(8) come from taking  measurements in real experiments.

\section{Circumventing the measurement issues of real experiments}

Hitherto, a technique has not been designed to easily, reliably and accurately measure some physical properties of the plasma in Z pinch discharges. This weakness shows up in plasma temperature measurement, for example. In contrast, other properties like the radius of the plasma column as a function of time can be easily measured. This fortunate circumstance allows us, in a way, to access data that is otherwise difficult to grasp. Indeed, if the dimensionless parameters $\alpha^2$ and $\beta$ associated with our experimental setup fall within the region where the snowplow model is valid or close to it, then we can trust Eqn.(14). Hence, to obtain the unknown dimensionless parameter $\gamma$, we have only to insert the experimentally collected $r(t)$ on the right hand side of Eqn.(15). Then, by substituting the resulting $F$ in Eqn.(16) we get $\gamma$.
In close connection with this, let us remember that in many cases there exists a nearly perfect parallel between the dynamics of the plasma sheath radius and that of the plasma column radius as they show up in snowplow model simulations and experiments, respectively.

Picking up the thread, once we know $\gamma$, we can directly access the internal energy of the plasma, $U$, using the formula

\begin{equation}
U=\gamma\,E_0.
\end{equation}
Additionally, the plasma temperature $k_BT$ can be obtained from the plasma internal energy through the formula

\begin{equation}
k_BT=\gamma\,\left(\frac{2}{3}\right)\left(\frac{E_0}{N_0}\right).
\end{equation}

\section{Results and Discussion}

It is important to highlight that the set of dimensionless parameters $(\alpha^2, \beta, \gamma)$ is all that is needed to provide a comprehensive overview of Z pinch discharges. Even more, $\alpha^2, \beta$ and $\gamma$ are related in the manner of

\begin{equation}
\gamma=f(\alpha^2, \beta).
\end{equation}
\\
But the lack of relevant experimental data still keeps the shape of the function $f$ a mystery. In contrast, Eqn.(16) provides a closed-form relationship among 
$\alpha^2, \beta$ and $\gamma$  but that result is nothing more than the solution of a mathematical model for Z pinch discharges. Hence, the relevance of Eqn.(16) will have to be judged, where possible, by comparing it with experiments.
In any case, as we mentioned before, measuring plasma temperature in Z pinch experiments is still an unsolved problem. Therefore, for the time being, using Eqn.(21) to estimate plasma temperature at the discharges seems to be a good starting point. 

\section{Conclusions}

Below, we highlight the substance of what this research work reveals to us.
\begin{enumerate}
\item The Z pinch problem can be described in terms of eight physical parameters.  Seven of these are control parameters while the eighth depends on those seven.
\item Alternatively, considering the postulates of dimensional analysis theory we came to the conclusion that the complete description of Z pinch experiments requires only three dimensionless parameters. Not only that, we went one step further and found out what those dimensionless parameters were.
\item So, the complete solution to the Z pinch problem could simply be represented as a surface in $3D$ space. We assign an axis to each of the dimensionless parameters of the problem.
\item Unfortunately, the surface in $3D$ space based exclusively on experimental measurements is not currently within our reach.
\item As a remedy to this situation, the snowplow model provides a closed-form solution to the Z pinch problem.  We take this result with caution because the snowplow model, like any other, is only a first approximation to understand the macroscopic physics of  real Z pinch experiments.
\item Luckily enough, an \textit{ad hoc} combination of the snowplow model and really feasible experimental measurements allows us to easily estimate the plasma temperature in any real Z pinch discharge.
\item In short, we have faced two possible scenarios: 
\\
i) We have enough experimental data to draw the surface in $3D$ space. Then, we do not need anything else; any forthcoming Z pinch experiment will be assigned a point on that surface.
\\
ii) We do not have enough experimental data to draw the surface in $3D$ space. Hence, we are constrained to navigate in the realm of modeling and approximations. There, we might consider the coarse-grained approach alluded in point 5 above. Alternatively, if what we want is to improve the estimates, then better we resort to the scheme of point 6 above. Whatever the case may be, we can also draw the surface in $3D$ space valid for all Z pinch experiments at once. It should naturally be clear that this surface in $3D$ space is only an approximation to the real surface in $3D$ space referred to in point 3 above.
\end{enumerate}

\end{document}